\documentstyle[12pt]{article}
\input epsf

\begin{document}

\begin{center}

{\bf Opinion evolution in closed community.}
\vspace{0.5cm}

Katarzyna Sznajd-Weron$^1$ and J\'ozef Sznajd$^2$
\vspace{0.5cm}

$^1$Institute of Theoretical Physics, University of Wroc{\l}aw,\\ 
pl. M. Borna 9, 50-204 Wroc{\l}aw, Poland\\
$^2$Institute of Low Temperature and Structure Research,\\ 
Polish Academy of Sciences, 50-950 Wroc{\l}aw, Poland

\end{center}

{\bf Abstract}
A simple Ising spin model which can describe a mechanism of making a decision
in a closed community is proposed. It is shown via standard Monte Carlo
simulations that very simple rules lead to rather complicated dynamics and to
a power law in the decision time distribution. It is found that a closed
community has to evolve either to a dictatorship or a stalemate
state (inability to take any common decision). A common decision can be
taken in a "democratic way" only by an open community.

\vspace{0.5cm}


\section{Introduction}
 The Ising spin system is undoubtedly one of the most
frequently used models of statistical mechanics. Recently, this
model has also become the most popular physics export article to
"other branches of science" such as biology, economy or sociology
\cite{Stauffer,Holyst,Derrida,FG}.
There are two main reasons for that: first verbalized by Nobel prize winner
Peter.B.Medawar - "physics envy" (quoted e.g. by R. Dawkins in \cite{Brockman})
- a syndrome appearing in some
researchers who would like to have such beautiful and relatively
simple models as physicists have (for example the Ising model). On the
other hand, some physicists would like to be better understandable
by non-physicists and create theories which could not be so
univocally verified or falsified like in the classical areas of
physics. It is rather obvious that Ising-type models cannot
explain origins of very complicated phenomena observed in complex
systems. However, it is believed that these kind of models could
describe some universal behaviour connected, for example, with
selforganization of systems \cite{Bak,BS} or in other words, justify the
existence of some power laws, which recently  makes many
scientists very happy \cite{Stauffer1}. 
Between many exotic applications of Ising spin models these referring 
to the problem of a democratic choice of a one among two possibilities 
seem to be the most natural \cite{Holyst,FG,Stauffer2,Galam,HKS}.

\section{Model}

Let us consider a community which time and again should
take a stand in some matter, for example vote on a president in 
a two-party system. If each member of the
community can take only two attitudes ($A$ or $B$) then in several
votes one expects some difference $m$ of voters for $A$ and against.
We assume three limiting cases:

(i)
all members of the community vote for $A$ (an "all $A$" state),

(ii)
all members of the community vote for $B$ (an "all $B$" state),

(iii)
50\% vote for $A$ and 50\% vote for $B$,

\noindent
which should be the stable solutions of our model.

The aim of the paper is to analyze the time evolution of $m$. To model
the above mentioned system we consider an Ising spins chain ($S_i;
i=1,2,\ldots N$) with the following dynamic rules:

-- if $S_iS_{i+1}=1$ then $S_{i-1}$ and $S_{i+2}$ take the
direction of the pair (i,i+1),  \hspace\fill  (r1)

-- if $S_iS_{i+1}=-1$ then $S_{i-1}$ takes the direction of
$S_{i+1}$ and $S_{i+2}$ the direction of $S_i$. \hspace\fill(r2)

These rules describe the influence of a given pair on the
decision of its nearest neighbours. When members of a pair
have the same opinion then their nearest neighbours agree with
them. On the contrary, when members of a pair have opinions different
then the nearest neighbour of each member disagrees with him (her). 
These dynamic rules lead to the three steady states above. 
However, the third steady state (50\% for $A$ and 50\% for $B$) 
is realized in a very special way.
Every member of the community disagrees with his (her) nearest neighbour
(it is easy to see that the Ising model with only next nearest
neighbour interaction has such fixed points: ferro- and antiferromagnetic
state). 
This rule is in accordance with the well known sentence 
"united we stand divided we fall ". 
So, from now on we will call our model - USDF.

\section{Isolated System}

To investigate our model we perform a standard Monte Carlo
simulation with random updating. 
We consider a chain of $N$ Ising spins with free
boundary conditions. In our simulations we were taking usually
$N=1000$, but we have done simulations also for $N=10000$. We
start from a totally random initial state i.e. to each site of the
chain we assign an arrow with a randomly chosen direction: up or
down (Ising spin). For this case we obtain of course, as a final
state, one of the three fixed points (1-3, i.e. AAAA, BBBB, ABAB) 
with probability 0.25,0.25 and 0.5, respectively. The typical relaxation 
time for $N=1000$ is $\sim 10^4$ Monte Carlo steps (MCS). The space
distribution of spins from the initial to a steady state is shown in
Fig. 1. For intermediate states one can see the formation of clusters.

\begin{figure}[htbp]
\centerline{\epsfxsize=12cm \epsfbox{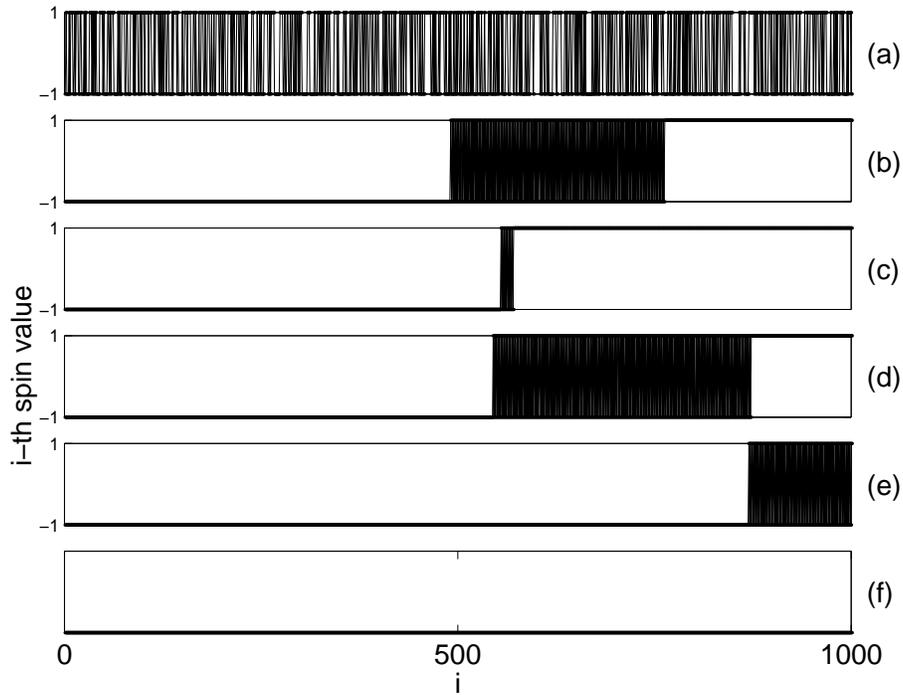}}
\caption{Spatial distribution of spins for (a) the initial state,
(b-e) interediate states, (f) the final (steady) state; time interval between
states is 10000 MCS}
\end{figure}

Let us define the decision as a magnetization, i.e.:
\begin{equation}
m=\frac{1}{N}\sum_{i=1}^N S_i.
\label{em}
\end{equation}

In Fig. 2 we present typical time evolution of $m$ and to compare
certain empirical data on "social mood" \cite{cbos}. 
Without any external stimulation decision can change
dramatically in a relatively short time. 
Such strongly non-monotonic behaviour of the change of $m$ 
is typically observed in the USDF model when the system evolves 
towards the third steady state (total disagreement or in magnetic 
language the antiferromagnetic state). 

\begin{figure}[htbp]
\centerline{\epsfxsize=12cm \epsfbox{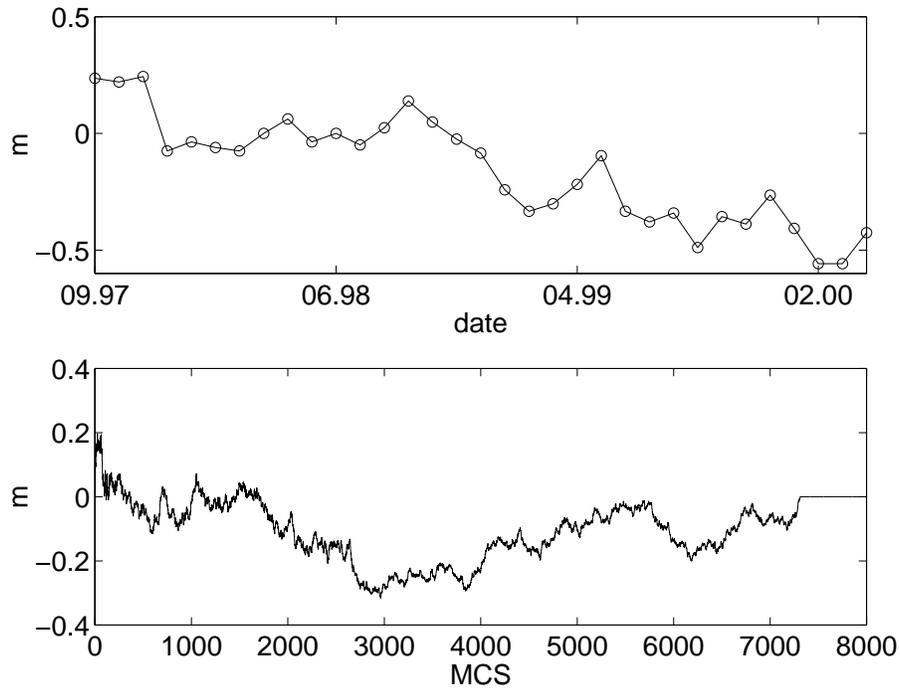}}
\caption{Time evolution of decision $m$ taken from empirical data 
\protect{\cite{cbos}}(upper) (question:"do you think future will be good?" asked
to $N=1100$ adults) and simulation 
from a random initial state (lower) for $N=1000$}
\end{figure}

To measure the time correlation of $m$ we use classical autocorrelation function:
\begin{equation}
G(\Delta t) = \frac
{\sum \left( m(t)-<m>\right) \left( m(t+ \Delta t)- <m>\right)} 
{\sum (m(t)-<m>)^2}.
\end{equation}
Comparison of simulation results with empirical data is shown in Fig.3.

\begin{figure}[htbp]
\centerline{\epsfxsize=12cm \epsfbox{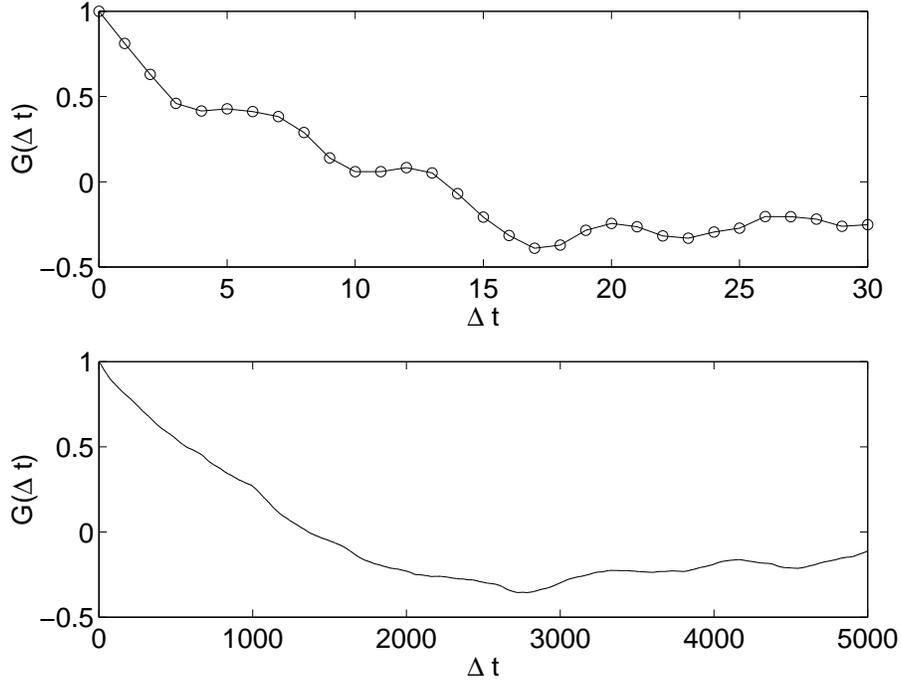}}
\caption{Autocorrelation for empirical data
(upper) and simulations (lower) evaluated from data shown on Fig.2
}
\end{figure}

It seems interesting to follow changes of one particular individual.
The dynamic rules we have introduced lead to an amazing
effect - if an individual changes his (her) opinion at time $t$ he
(she) will probably
change it also at time $t+1$. Like in the Bak Sand-Pile
model one change can cause an avalanche \cite{Bak,BS}. On the other hand
an individual can stay for a long time without changing his (her) decision.
Let us denote by $\tau$ the time needed by an individual to change  his (her)
opinion. From Fig. 3 it can be seen that $\tau$ is usually very short,
but sometimes can be very long. The distribution of $(\tau)$ ($P(\tau)$) 
follows a power law with an exponent $-3/2$ (see Fig. 4).

\begin{figure}[htbp]
\centerline{\epsfxsize=12cm \epsfbox{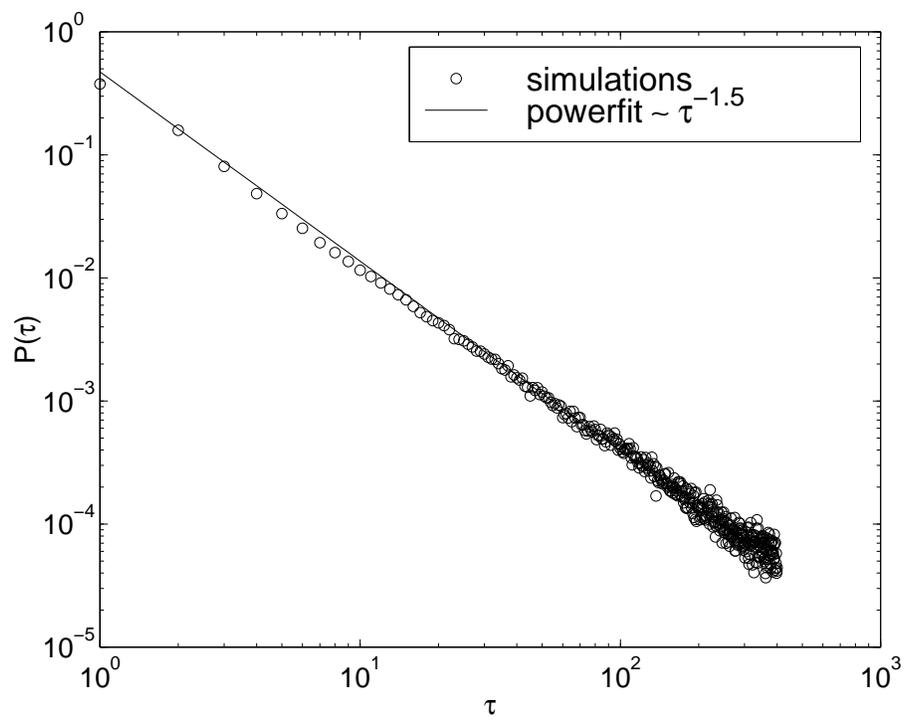}}
\caption{Distribution of decision time $\tau$ follows a power law.
}
\end{figure}

We have also analysed the influence of the initial conditions on the
evolution of the system.
We have done it in two different ways - randomly and in clusters.
In both cases we start from an initial concentration $c_B$ of opinion $B$.
In the random setup $c_B*N$ individuals are randomly (uniformly) chosen
out of all $N$ individuals. In the cluster setup simply the first $c_B*N$
individuals are chosen. 

It turns out that the distribution of decision time $\tau$ still follows 
the power law with the same exponent as shown in Fig.4. A non-monotonic 
behaviour of decision change is still typical and sometimes even much stronger. 
However, it is obvious that if initially there is more $A$'s then $B$'s 
the final state should be more often "all $A$" then "all $B$". Dependence
between initial concentration of $B$ and the probability of steady
state $S$ ($AAAA$,$BBBB$ or $ABAB$) is shown in Fig. 5. 

\begin{figure}[htbp]
\centerline{\epsfxsize=12cm \epsfbox{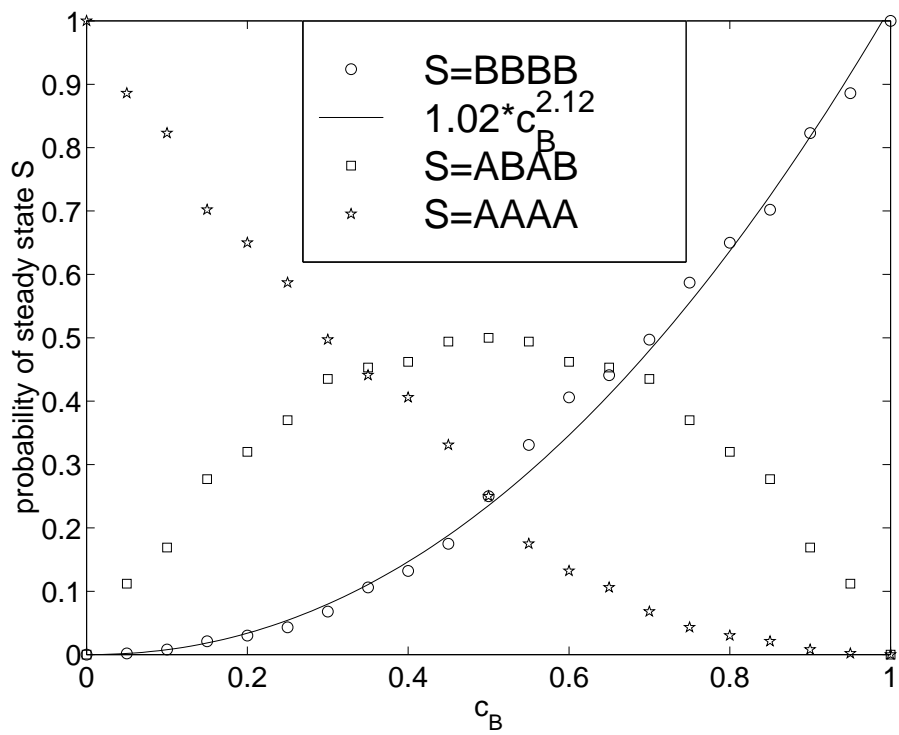}}
\caption{Dependence between the initial concentration of $B$ and the final
state. Averaging was done over 1000 samples.
}
\end{figure}

There is no significant difference between the "cluster" and "random case".
Although it should be noted that the "random case"
is not well defined, because there is a number of random
initial states. However, averaging over different initial random conditions
gives a similar result to the "cluster case".

Observe that $c_B>0.7$ is needed to obtain final state "all B" with
probability grater than $0.5$ (see Fig.5). Dependence between $c_B$ and the
probability of steady state "all B" is well fitted by a power function with
an exponent $2.12$.

\section{Information noise}

It is well known that the changes of opinion are determined by the 
{\it social impact} \cite{Latane}.
In the previous section we have considered a community in which
a change of an individuals opinion is caused only by a contact
with its neighbours. It was the simplest {\it social impact} one can imagine. 
Now, we introduce to our model noise $p$ (similar to the
"social temperature" \cite{Holyst}), which
is the probability that an individual, instead of following the
dynamic rules, will make a random decision. We start from an "all $A$" 
state to investigate if there is a $p \in (0,1)$ which does not throw the
system out of this state. Time evolution of the decision from the "all
A" state is shown in Fig. 6.

\begin{figure}[htbp]
\centerline{\epsfxsize=12cm \epsfbox{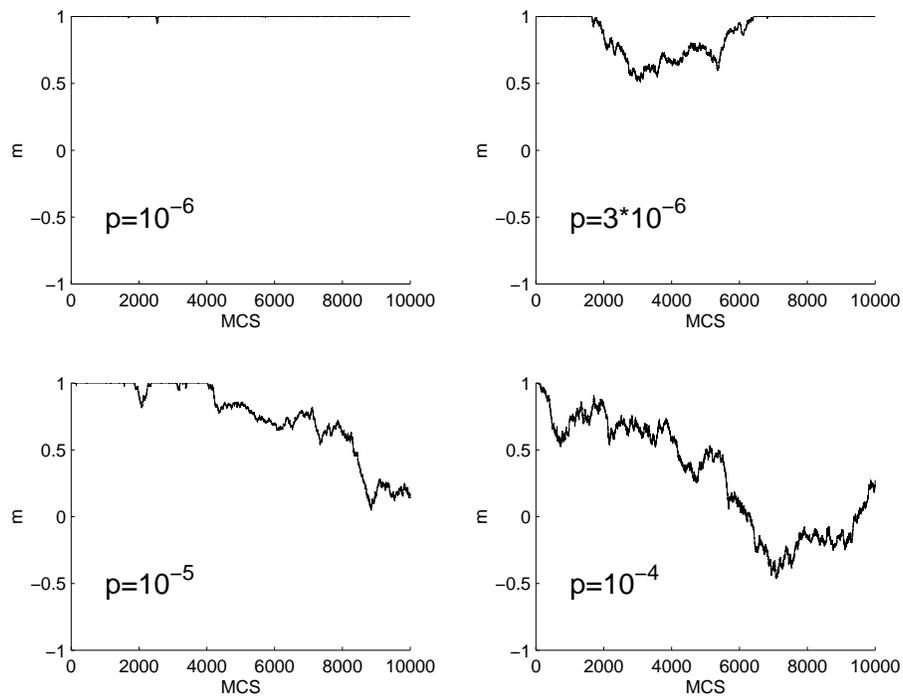}}
\caption{Figure 6: Time evolution of decision $m$ from an initial state "all A" for different values of noise $p$.}
\end{figure}

It can be seen that for very small $p
\sim 10^{-6}$ deviations from the steady state are slight and the
system is almost totally ordered. If $p$ increases than the
deviations are of course larger and the system goes to a completely
disordered state. This suggests that there is some value of $p =
p^*$ below which the system is ordered. However, on the basis
of our simulations we can only determine that $p^* \in [10^{-6},
10^{-5}]$. If we take totally random initial conditions and
$p<p^*$ the system will reach, after some relaxation time, one of
the three steady states. 

In the previous section we have
shown that the distribution $P(\tau)$ follows a power law with an
exponent $\sim -3/2$. The same distribution for different values
of $p$ is shown in Fig. 7. In the limit $p \rightarrow 0$
distribution $P(\tau)$ indeed follows a power law, whereas for
$p \rightarrow 1$ the distribution is exponential.
Between these two extreme values of $p$ the distribution $P(\tau)$
consists of two parts - exponential for large  values of $\tau$ and
a power law with exponent $\omega \sim -3/2$ for small
values of $\tau$. Thus we can write:
\begin{equation}
P(\tau) \sim
\left \{
\begin{array}{ll}
\tau^{-\omega} & \mbox{for $\tau < \tau^*$},\quad \omega \sim
\frac{3}{2}\\ \exp(a\tau) & \mbox{for $\tau > \tau^*$},
\end{array}
\right.
\end{equation}
where $\tau^*$ decreases with increasing $p$. For $p=0$, $\tau^*=\infty$
and for $p=1$, $\tau^*=0$.

\begin{figure}[htbp]
\centerline{\epsfxsize=12cm \epsfbox{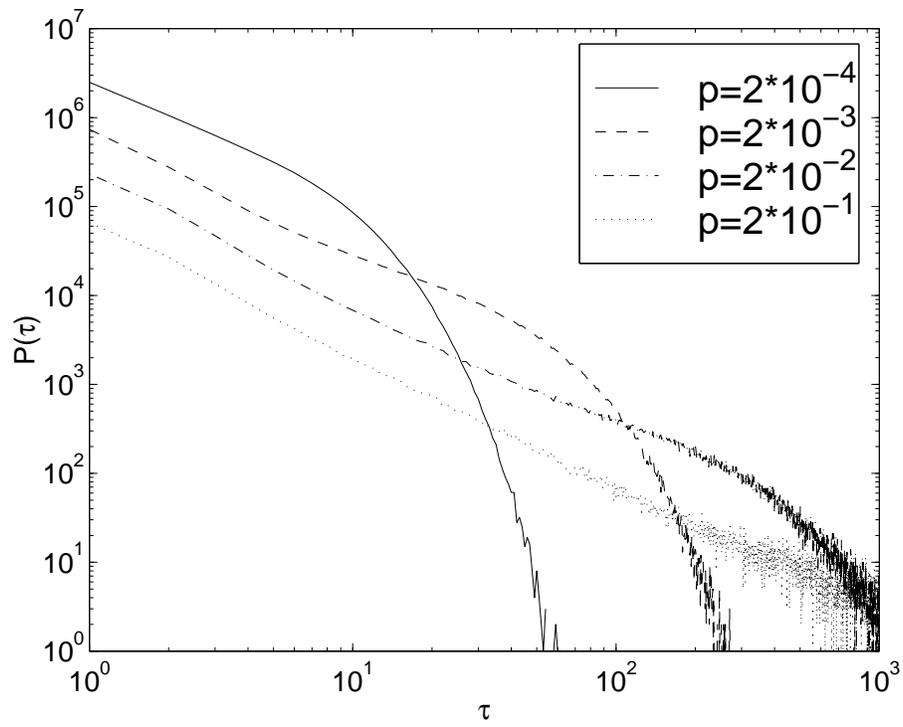}}
\caption{Distribution of decision time $P(\tau)$ for different
values of noise $p$.
}
\end{figure}

\section{Discussion}

We have proposed a simple model called USDF, which can describe a
"black and white" way of making a decision, where unanimity or
disagreement of a given pair causes unanimity or disagreement to its
nearest neighbours, respectively. If such a mechanism of taking a
decision by a community is correct our model leads to the
following conclusions:

(i)
In a closed (isolated) community there are only two possibilities
of a final state: dictatorship or stalemate. After a shorter or
longer time our model tends to one of the steady "ordered" states
(1-3). It means either a total unanimity if the system goes to the
state 1(2) or inability to take any common decision if it goes to
3. However, small but finite information noise (open
community) leads to disorder and the system does not go to any
steady state. In this case there is a possibility of taking a common
decision in a democratic way.

(ii)
A change of opinion is followed by further changes. 
Periods of frequent changes of opinions are followed by
periods of stagnancy.

(iii)
A relatively small group (a few percent of the whole population)
by a favourable coincidence can bring to a stalemate. 
But in order to win the group has to be quite large. For example,
if the group wants to have a 50\% chance of winning it should
consist of more than 70\% of all individuals. 
It means that in order to change an existing law 
(for example that pornography is illegal)
usually over 70\% of the population have to vote for change.
A similar effect was observed by Galam \cite{Galam}.

(iv)
For finite information noise p, there is some characteristic
time of a decision change $\tau*$ which depends on a value of p.
For the decision time less than  $\tau*$, the distribution of
decision time $\tau$ follows a power-law.

(v)
The distribution of decision time $\tau$ for $p=0$ and for 
$p<1$ and $\tau<\tau*$ follows the "universal" power-law with 
exponent $\omega \sim 3/2$ independently of the initial conditions
(totally random or clustered state).

The proposed very simple rules (r1, r2) leads to a rather complicated
dynamics but one can doubt if these rules properly describe real
mechanisms of taking a decision. There are of course other
possibilities within the Ising spin model. For example, if the
members of a given community are less prone to oppose nearest
neighbours then one should keep rule (r1) and skip rule (r2). 
This means that if $S_{i}S_{i+1}=-1$ then nothing is changed
in the system. In this case there are only two steady states "all
A" and "all B". Our simulations suggest that in such a closed
community there is a tendency to create two opposite clusters but
the final state must be total unanimity (dictatorship). This
result is rather obvious, it is easier to carry one's opinion if
the members of the community are peaceable or less active. For a
while this can be of profit to the community but finally it must
lead to a dictatorship. It should be noted that also in this case
the distribution of the decision time follows the same power-law
like in the USDF model with $\omega \sim 3/2$.

In an other model which we call "if you do not know what to do, just
do nothing" an individual's opinion depends on opinions of his (her)
nearest neighbours. If the opinion of these neighbours is
unanimity ($S_{i-1}S_{i+1}=1$) then the i-th individual agrees
with them if not the i-th individual does not change his (her) opinion. In
this case the number of steady states is enormous. Namely, each
state different than $ABAB$ (antiferromagnetic) is a steady
state. It means that if we start with a random distribution of
opinions A and B there will be a tendency to create small clusters.

It is rather obvious that in a real community all mentioned
and many others mechanisms can effect an opinion evolution...

{\bf Acknoledgements}

We are grateful to Professor Suzanna Moss de Oliveira 
and Professor Dietrich Stauffer for reading the paper and helpful
remarks.

\end{document}